\documentclass{pasj00}

\usepackage{color}
\usepackage{ulem}

\Received{yyyy/mm/dd}
\Accepted{yyyy/mm/dd}
\Published{$\langle$publication date$\rangle$}
\SetRunningHead{S.\ Takeuchi, K.\ Ohsuga, and S. Mineshige}{
Clumpy Outflow Formation
}

\usepackage{times,bm}

\begin{document}

\title{Clumpy Outflow from Supercritical Accretion Flows}
\author{Shun \textsc{Takeuchi},\altaffilmark{1}\thanks{Present address:  Fujitsu Limited, 1-9-3 Nakase, Mihama-ku, Chiba-shi, Chiba 261-8588, Japan}
       Ken \textsc{Ohsuga},\altaffilmark{2, 3}
        and
        Shin \textsc{Mineshige}\altaffilmark{1}
        }
\altaffiltext{1}{Department of Astronomy, Graduate School of Science, Kyoto University, Sakyo-ku, Kyoto 606-8502, Japan}
\altaffiltext{2}{National Astronomical Observatory of Japan, Osawa, Mitaka, Tokyo 181-8588, Japan}
\altaffiltext{3}{School of Physical Sciences, Graduate University of Advanced Study (SOKENDAI), Shonan Village, Hayama, Kanagawa 240-0193, Japan}
\email{E-mail : shun@kusastro.kyoto-u.ac.jp}

\maketitle

\begin{abstract}
Significant fraction of matter in supercritical (or super-Eddington) 
accretion flow is blown away by radiation force, thus forming outflows, however, the properties of such radiation-driven outflows have been poorly understood.
We have performed global two-dimensional radiaion-magnetohydrodynamic simulations 
of supercritical accretion flow onto a black hole with 10 or $10^8 M_{\odot}$ in a large simulation box 
of $514 r_{\rm S} \times 514 r_{\rm S}$ (with $r_{\rm S}$ being the Schwarzschild radius). 
We confirm that uncollimated outflows with velocities of 10 percents of the speed of light 
emerge from the innermost part of the accretion flow 
over wide angles of $10^{\circ}$--$50^{\circ}$ from the disk rotation axis. Importantly, 
the outflows exhibit clumpy structure above heights of $\sim 250 r_{\rm S}$. 
The typical size of the clumps is $\sim 10 r_{\rm S}$, 
which corresponds to one optical depth, and their shapes are slightly elongated along the outflow direction. 
Since clumps start to form in the layer above which (upward) radiation force overcomes (downward) gravity force, 
Rayleigh-Taylor instability seems to be of primary cause.
In addition, a radiation hydrodynamic instability,
which arises when radiation funnels through radiation-pressure supported atmosphere,
may also help forming clumps of one optical depth.
Magnetic photon bubble instability seems not to be essential, 
since similar clumpy outflow structure is obtained in 
non-magnetic radiation-hydrodynamic simulations.
Since the spatial covering factor of the clumps is estimated to be $\sim 0.3$ 
and since they are marginally optically thick, 
they will explain at least some of rapid light variations of active galactic nuclei.
We further discuss a possibility of producing broad-line clouds by the clumpy outflow.
\end{abstract}

\section{Introduction}
There is growing observational evidence supporting that significant amount of material
is blown away from accretion flow (or disk) onto black holes in forms of outflows and jets.
In most cases, the outflows are observed through blue-shifted absorption line features
(e.g., \cite{Mil+04}; \cite{Kot+06}; \cite{Kub+07}; \cite{Nei+12}; \cite{Pon+12} for Galactic sources
and \cite{TerWil01}; \cite{Pou+03}; \cite{Ree+03}; \cite{GanBro08}; \cite{PouRee09} 
for active galactic nuclei: AGNs).
The estimated mass outflow rate sometimes exceeds the mass accretion rate to the central black holes 
(e.g.,  \cite{Mil+08}; \cite{Ued+09}; \cite{Nei+11}).
The dense outflows can also produce Compton upscattering soft photons 
from the underlying accretion flow, thereby making the emergent spectra harder 
(e.g, \cite{Gla+09}; \cite{Kaw+12}).
The structure of the accretion flow is also affected by the presence of outflows
(\cite{KawMin09}).

In addition, the outflows are expected to produce deep impacts on their environments,
since the outflows from black hole accretion flow,
especially those from the innermost part of the flow,
do carry large amount of mass, momentum, and energy.
Such powerful outflows may cause dynamical feedback which may trigger (or quench)
star formation in surrounding interstellar and galactic medium, thus promoting
metal enrichment (\cite{Mur+95}; \cite{Gra+04}; \cite{DiM+05}).
Larger impacts are expected by higher luminosity objects
with larger mass accretion rates.

At luminosities close to the Eddington luminosity,
radiation energy density overcomes matter energy density 
so that radiation field can also be dynamically important;
radiation-driven outflow thus naturally arises.
Matter emits, absorbs, and scatters radiation, 
while radiation gives (or removes) energy and momentum to matter.
In this way, matter and radiation are strongly coupled to each other.
To study the outflow from luminous systems,
we thus need to properly simulate strong coupling between matter and radiation.

There are two basic lines of simulation study of outflow from
luminous black hole accretion flow: hydrodynamic simulations of line-driven outflow
by Proga and collaborators (\cite{Pro+98}; \cite{Pro+00}),
and radiation-hydrodynamic simulations of continuum-driven outflow
by our group (\cite{Ohs+05}; \cite{Tak+09}).
By performing the hydrodynamic simulations incorporating 
radiation force due to spectral lines (line force) into the equation of motion, 
\citet{ProKal04} have studied the dynamics of the line-driven outflow.
They investigated the outflow properties above the disk
by setting an optically thick and geometrically thin disk on the equatorial plane as a boundary condition.
The radiation energy equation is not solved and the reprocessed photons are neglected.
In contrast, our simulations self-consistently solve the accretion disks and the outflows, 
although our studies are restricted to the continuum-driven outflows.
By solving radiation energy equation,
we consider diffusive photons which suffer numerous Thomson scattering.
The continuum radiation force via diffusive photons drives outflows.

In traditional hydrodynamic (or radiation-hydrodynamic) models of accretion
flows, the disk viscosity, most important physical process in the disk theory,
is described by the phenomenological $\alpha$-viscosity model,
although it should, in principle, be determined in accordance with
magnetic field amplifications and dissipation within the flow.
In other words, global multi-dimensional radiaion-magnetohydrodynamic (radiation-MHD)
simulations are necessary for understanding luminous black hole inflow-outflow system.
Under such considerations, we have recently started global two-dimensional radiation-MHD simulations
(\cite{Ohs+09}; \cite{OhsMin11}, hereafter Papers I and II, respectively).
The most important findings in those simulations are ubiquitous outflow
in every mode of accretion flow
and a new type of jet emerging at high luminosities;
magnetically collimated, radiation-pressure driven jet (\cite{Tak+10}).

The aim of the present study is to examine large-scale behaviour of
continuum-driven outflow from supercritical (or super-Eddington) accretion flows. For this purpose,
we run new simulation with a significantly increased simulation box size of
$514 r_{\rm S}\times 514 r_{\rm S}$,
where $r_{\rm S}$ is the Schwarzschild radius
(Note that it was only $105 r_{\rm S}\times 103 r_{\rm S}$  in Papers II).
With this treatment we could find clumpy outflow structure
in distant regions from the central black hole.
The plan of this paper is as follows. In the next section, 
we overview the basic equations of our radiation-MHD simulations.
We present the simulation results of clumpy outflow from supercritical accretion flows in section 3.
The formation mechanisms of clumpy structure and 
the observational implications will be also discussed in section 4.
The final section is devoted to concluding remarks.

\section{Basic Equations}
We simulated supercritical accretion flows and the associated outflows
by using the global two-dimensional radiation-MHD code developed by Papers I and II. 
The basic equations and the numerical method are described them in details.
Hence, we overview them and stress the differences from the previous calculations.

We used cylindrical coordinates $(R,\theta, z)$,
where $R$, $\theta$, and $z$ are the radial distance, the azimuthal angle,
and the vertical distance, respectively.
We assume that the flow is non-self-gravitating,
reflection symmetric relative to the equatorial plane (with $z = 0$),
and axisymmetric with respect to the rotation axis (i.e., ${\partial}/{\partial \theta} = 0$).
The basic equations which take the terms up to the order of $(v/c)^1$ are as follows
(\cite{MihMih84}; \cite{StoNor92}):
the continuity equation,
\begin{equation}
\frac{\partial \rho}{\partial t}
+ {\bm \nabla} \cdot (\rho {\bm v}) = 0,
\label{mass_con}
\end{equation}
the equations of motion,
\begin{eqnarray}
\frac{\partial (\rho {\bm v})}{\partial t} + {\bm \nabla} \cdot
\left( \rho {\bm v}{\bm v} - \frac{{\bm B}{\bm B}}{4 \pi} \right)
 = &-& {\bm \nabla} \left( p_{\rm gas}+\frac{B^2}{8\pi} \right) \nonumber \\
&+& \frac {\chi}{c} {\bm F}_0 -\rho{\bm \nabla}\psi_{\rm PN},
\label{mom}
\end{eqnarray}
the energy equation of the gas,
\begin{eqnarray}
\frac{\partial e }{\partial t}
+ {\bm \nabla}\cdot(e {\bm v}) 
 = &-& p_{\rm gas}{\bm \nabla}\cdot{\bm v} +\frac{4\pi}{c^2}\eta J^2 \nonumber \\
 &-& 4\pi \chi_{\rm abs} B_{\rm bb} + c \chi_{\rm abs} E_0,
\label{gase}
\end{eqnarray}
the energy equation of the radiation,
\begin{eqnarray}
\frac{\partial E_0}{\partial t} + {\bm \nabla}\cdot(E_0 {\bm v}) 
 = &-& {\bm \nabla} \cdot {\bm F_0} - {\bm \nabla} {\bm v}:{{\bf P}_0} \nonumber \\
&+& 4\pi \chi_{\rm abs} B_{\rm bb} - c \chi_{\rm abs} E_0,
\label{rade}
\end{eqnarray}
and the induction equation,
\begin{equation}
\frac{\partial B}{\partial t}
={\bm \nabla} \times
\left({\bm v}\times {\bm B}-\frac{4\pi}{c}\eta {\bm J} \right).
\label{ind}
\end{equation}
Here, $\rho$ is the matter density,
$\bm{v}$ is the flow velocity,
$c$ is the speed of light,
$e$ is the internal energy density of the gas,
$p_{\rm gas}$ is the gas pressure,
$\bm{B}$ is the magnetic field,
$\bm{J} = c{\bm \nabla}\times {\bm B}/4\pi$
is the electric current,
$B_{\rm bb} =\sigma T_{\rm gas}^4/\pi$ is the blackbody intensity,
$\sigma$ is the Stefan-Boltzmann constant,
$T_{\rm gas}$ is the temperature of the gas,
$E_0$ is the radiation energy density,
${\bm F}_0$ is the radiative flux,
${ {\bf P}}_0$ is the radiation-pressure tensor,
$\chi_{\rm abs} = (\kappa_{\rm ff} + \kappa_{\rm bf}) \rho$ is the absorption opacity
per unit volume (with dimensions of length$^{-1}$),
$\chi = (\kappa_{\rm es} + \kappa_{\rm ff} + \kappa_{\rm bf}) \rho$ is the total opacity per unit volume,
and $\eta$ is the resistivity, respectively.
The subscript $0$ means the values measured in the co-moving (fluid) frame.
For simplicity, we adopt the gray (frequency-integrated) approximation for the radiation terms.

We adopt the pseudo-Newtonian potential, $\psi_{\rm PN}$,
to incorporate the general relativistic effects,
given by $\psi_{\rm PN} = -GM/(r-r_{\rm S})$ (\cite{PacWii80}).
Here, $r = (R^2+z^2)^{1/2}$ is the distance from the origin.
The Schwarzschild radius is given by $r_{\rm S} =2GM/c^2$, 
where $G$ is the gravitational constant,
and $M$ is the mass of the black hole, respectively.

The set of equations (\ref{mass_con})-(\ref{ind}) can be closed by using an
ideal gas equation of state,
$p_{\rm gas} = (\gamma -1)e = \rho k_{\rm B} T_{\rm gas}/\mu m_{\rm p}$,
and by adopting the flux-limited diffusion (FLD) approximation
to evaluate $\bm F_0$ and ${\bf P}_0$ (\cite{LevPom81}).
Here, $\gamma$ is the specific heat ratio, $k_{\rm B}$ is the Boltzmann constant,
$\mu$ is the mean molecular weight, and  $m_{\rm p}$ is the proton mass.

We consider the electron scattering, $\kappa_{\rm es} $, the Rosseland mean
free-free absorption, $\kappa_{\rm ff}$, and bound-free absorption opacity, $\kappa_{\rm bf}$,
for solar metallicity;
\begin{equation}
\kappa_{\rm es} = \sigma_{\rm T} m_{\rm p}^{-1},
\end{equation}
\begin{equation}
\kappa_{\rm ff} = 1.7 \times 10^{-25} m_{\rm p}^{-2} \rho T_{\rm gas}^{-7/2} \; {\rm cm}^2{\rm g}^{-1},
\end{equation}
and
\begin{equation}
\kappa_{\rm bf} = 4.8 \times 10^{-24} m_{\rm p}^{-2} \rho T_{\rm gas}^{-7/2}  \; {\rm cm}^2{\rm g}^{-1},
\end{equation}
where $\sigma_{\rm T}$ is the Thomson scattering cross-section.

We initially set a magnetized rotating torus surrounding a non-rotating black hole at the origin.
The center of the torus, where the matter density is at maximum, is located at $R = 40 \, r_{\rm S}$.
The initial, maximum density (at the center of the torus) is the same
as that of Model A of Paper I and II; i.e., $\rho_0 = 1$ g cm$^{-3}$.
We assume initially closed poloidal magnetic field
and their strengths are determined so as to achieve the plasma-$\beta$
$(\equiv p_{\rm gas}/p_{\rm mag}$) to be $100$ within the torus,
where $p_{\rm mag} = B^2/8\pi$ is the magnetic pressure.
This torus is sandwiched by a non-rotating, non-magnetized isothermal corona.
Since the initial ambient gas is finally ejected out of the computational domain,
it does not affect the resulting flow structure.

We adopt free boundary conditions at the upper and the outer boundaries,
where we allow mass, momentum, and energy to free go out but
nothing can enter the calculation box.
At the inner boundary at $R = 2 r_{\rm S}$ around the black hole
we remove all the mass, momentum, and energy
which enter the inner zone inside this radius.
Technically, we first solve the non-radiative MHD equations for initial 1 s, 
and then turn on the radiation terms.

The calculation methods, the initial model, and the boundary conditions so far described are
 the same as those adopted in Papers I and II.
Only a difference is the size of the calculation box.
Since the purpose of this study is to examine global gas dynamics of the outflow
originating from supercritical accretion flow, we need
a wider spatial range.
Thus, we set the computational domain of cylindrical shells of
$2 \,r_{\rm S} \leq R \leq 514 \,r_{\rm S}$ and
$0 \leq z \leq 514 \,r_{\rm S}$.
This large calculation box leads us to a discover of clumpy outflow structure,
as we see in the next section.
The grid spacing of the radial distance, $\Delta R$, and the vertical distance,
$\Delta z$, are uniform, given by $\Delta R = \Delta z = 0.4 \, r_{\rm S}$.
Note that the computational domain of Papers II
was $2 \,r_{\rm S} \leq R \leq 105 \,r_{\rm S}$,
and $0 \leq z \leq 103 \,r_{\rm S}$,
and the grid spacing was $\Delta R = \Delta z = 0.2 \, r_{\rm S}$.

\section{Properties of Clumpy Outflow}
\subsection{Overview of Clumpy Outflow}
\begin{figure*}
\begin{center}
\FigureFile(157mm,110mm){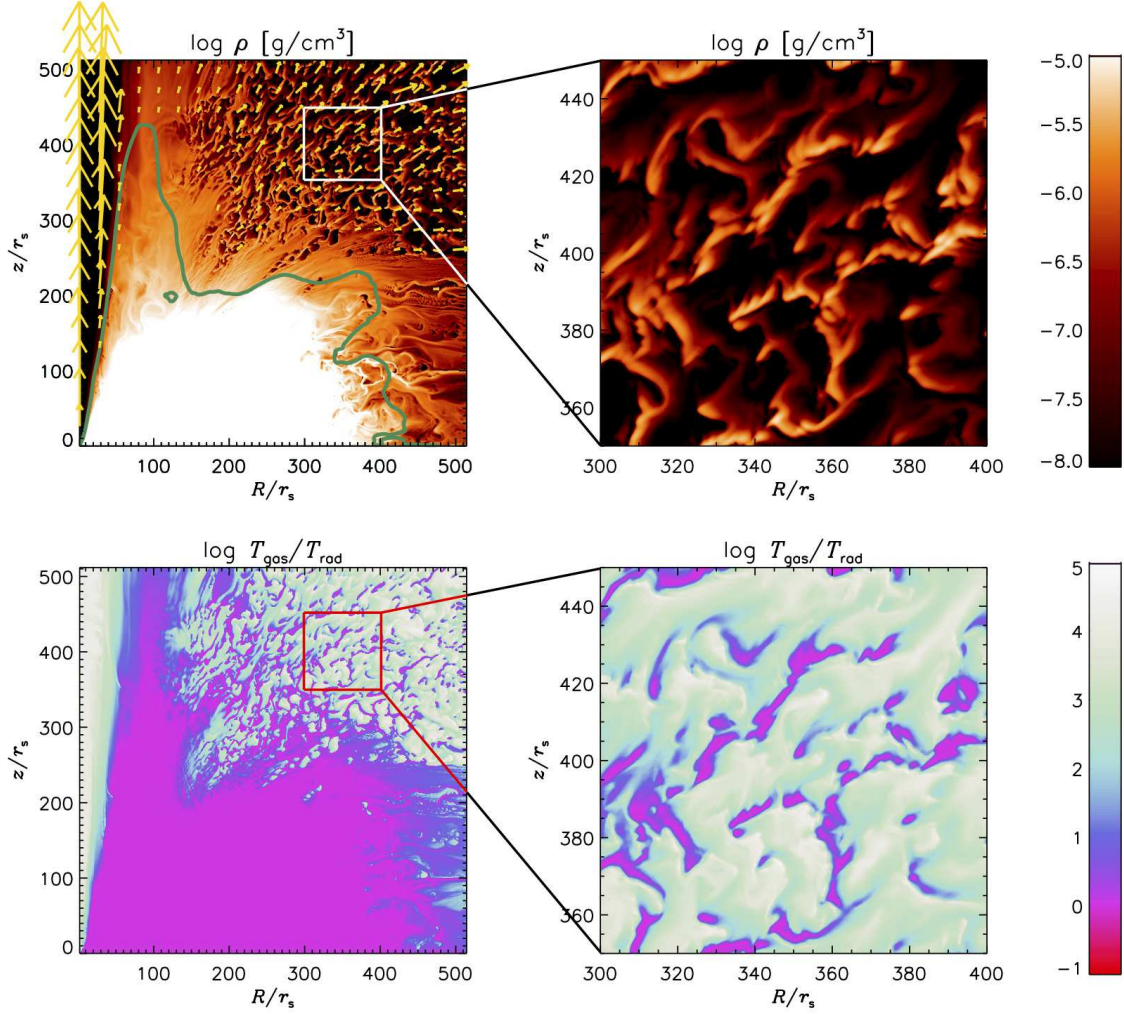}
\end{center}
\caption{
Two-dimensional structure of a supercritical accretion flow and the associated outflow
around a black hole with $M = 10 M_{\odot}$.
The upper and lower left panels, respectively, show the color contours of the matter density 
overlaid with the flow velocity vector and 
those of the ratio of the gas temperature to the radiation temperature
in the whole computational domain.
The velocity arrows are displayed only in the region  in which their values exceed the escape velocity.
Green lines in the upper left panel indicate the surface which upward radiation force equals
downward gravity force of the central black hole.
The right two panels are the same as those in the left panels but in a narrower region;
$300 r_{\rm S} \le R \le 400 r_{\rm S}$ and $350 r_{\rm S} \le z \le 450 r_{\rm S}$.
The elapsed time is $9$ s, which corresponds to about 30 times Keplerian timescale at $R = 40 r_{\rm S}$.
}
\label{fig:rho_contour}
\end{figure*}

Let us first overview the global gas dynamics.
Figure~\ref{fig:rho_contour}
shows the two-dimensional distributions of the matter density (upper panels)
and the ratio of the gas temperature to the radiation temperature (lower panels)
in the whole computational domain in the left panels 
and in a clumpy outflow region (defined later) in the right panels 
at the elapsed time of $t = 9$ s.
The black hole mass is set to be $M = 10 M_\odot$.
The radiation temperature is given by $T_{\rm rad} = (c E_{0}/4\sigma)^{1/4}$.
Yellow arrows in the upper left panel indicate the velocity field
whose values are larger than the escape velocity, $v_{\rm esc} = (2GM/r)^{1/2}$.
Green lines in the upper left panel indicate the surface,
where the equality, $\chi F_0/c =  \rho |\nabla \psi_{\rm PN}|$, holds in time average between $t = 8.8$--$9.2$ s.
Upward radiation force overcomes
downward gravity force of the central black hole in the upper and right region.

We evaluated the following basic quantities:
the mass accretion rate is $\dot{M}_{\rm acc} \sim 100 L_{\rm E}/c^2$;
the mass outflow rate is $\dot{M}_{\rm out} \sim 10 L_{\rm E}/c^2$;
the photon luminosity is $L \sim L_{\rm E}$.
Here, the Eddington luminosity $L_{\rm E}$ is given by $L_{\rm E} = 4\pi c G M/\kappa_{\rm es}$.
The mass accretion rate and the mass outflow rate were calculated by
summing up the mass passing through the inner boundary
and the upper boundary per unit time with higher velocities
than the escape velocity. 
The photon luminosity were calculated based
on the radiative flux at the upper boundary:
\begin{equation}
\dot{M}_{\rm acc} = - 2 \int^{2r_{\rm S}}_{0} 2 \pi R \rho v_R dz,
\end{equation}
\begin{equation}
\dot{M}_{\rm out} = 2 \int^{514 r_{\rm S}}_{2 r_{\rm S}} 2 \pi R \rho v_z dR,
\end{equation}
and
\begin{equation}
L = 2 \int^{514 r_{\rm S}}_{2 r_{\rm S}} 2 \pi R \left(  F^z_0 + v_z E_0   \right) dR.
\end{equation}
The mass accretion rate onto the black hole 
exceeds the critical rate giving rise to the Eddington luminosity;
i.e., the simulated flow is supercritical
(\cite{ShaSun73}; see chap. 10 of \cite{Kat+08} for a review).
The inflow-ouflow structure in the central region
($R \ltsim 100 r_{\rm S}$, $z \ltsim 100 r_{\rm S}$)
is consistent with the previous result (Paper II).
Note that  we adjusted the color scale in such a way to clearly show
the flow structure in the distant region from the central black hole.
As a result, the color scale in the central accretion flow region is saturated.

Figure 1 shows clumpy structure in the distant outflow region
($R \gtsim 200 r_{\rm S}$ and $z \gtsim 250 r_{\rm S}$),
in which upward radiation force overcomes downward gravity force.
Gas clumps whose matter density is about $10^{-6}$ g cm$^{-3}$
and velocity is about $10\%$ of the speed of light
are blown away over wide angles of $10^\circ$--$50^\circ$
from the disk rotation axis.

Besides the uncollimated outflows
we also confirmed the ejection of a high-speed, collimated jet
in a narrow region along the disk rotation axis.
This high-speed outflow is accelerated by continuum radiation force
while it is collimated by the Lorentz force of a magnetic tower structure
despite radiation energy greatly dominating the magnetic energy (see \cite{Tak+10}).
Note that the magnetic tower structure is created by the inflation
of toroidal magnetic field which are accumulating around the black hole.
In the present paper, we focus properties of the clumpy outflow.

We show the clumpy structure in more details in
the upper right panel of Figure~\ref{fig:rho_contour}.
The depicted area is $300 r_{\rm S} \le R \le 400 r_{\rm S}$ 
and $350 r_{\rm S} \le z \le 450 r_{\rm S}$
(hereafter called as the {\lq\lq}clumpy outflow region{\rq\rq}),
corresponding to the location of the white square in the upper left panel.
Clearly displayed is grossly inhomogeneous density pattern
which is composed of high density clumps ($\rho_{\rm cl} \sim 10^{-6}$ g cm$^{-3}$)
and ambient low-density media ($\rho_{\rm amb} \sim 10^{-8}$ g cm$^{-3}$).

The lower right panel shows the ratio of the gas temperature to the radiation temperature
in the clumpy outflow region.
It is interesting to note that the radiative equilibrium 
($Q_{\rm rad}^{+} \sim Q_{\rm rad}^{-}$, and hence $T_{\rm gas} \sim T_{\rm rad}$) is achieved within clumps, 
as well as within the underlying accretion flow.
Here, $Q^{+}_{\rm rad} = c \chi_{\rm abs} E_0$ and
$Q^{-}_{\rm rad} = 4\pi \chi_{\rm abs} B_{\rm bb}$ are
the radiation heating rate and the radiation cooling rate, respectively.
In the ambient media, in contrast, decoupling of matter and radiation occurs
($T_{\rm gas} > T_{\rm rad}$) because of low matter density.

Also note that Figure~\ref{fig:rho_contour} gives a snapshot.
In fact, the clumpy structure is not stationary but changes its shape in time.
The gas particles that compose a clump are not the same.  
As we will see later the gas particles move faster than
clumps and sometimes go through a clump.

\subsection{Clump size and shape}
The two-dimensional contour plots shown in the previous section are useful 
to understand the clumpy patterns of the outflow, but it is not always easy to derive
a typical size of the clumps and a mean separation between the clumps.
For deriving such statistical quantities, auto-correlation function (ACF) is quite useful
(see Appendix for the actual calculation method).

\begin{figure}
\begin{center}
\FigureFile(85mm,55mm){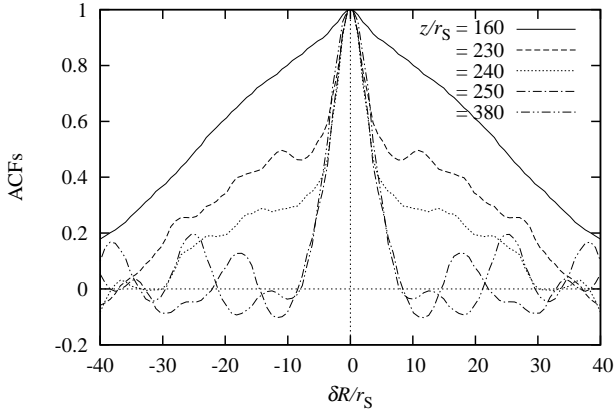}
\end{center}
\caption{
Auto-correlation functions (ACFs) of the matter density  in the $R$-direction ($\delta R$)
as a function of the typical height at the elapsed time of $t = 9$ s.
The width of the central peak represents the typical clump size,
whereas the separation to its neighbouring peak represents the typical clump separation.
}
\label{fig:auto}
\end{figure}

We show ACFs of the matter density as a function of the typical height 
at the elapsed time of $t = 9$ s in Figure~\ref{fig:auto}.
The integration range of the ACFs are 
$(R_{\rm in}, R_{\rm out}) = (100 r_{\rm S}, 200 r_{\rm S})$ for $z = 160 r_{\rm S}$, 
$(R_{\rm in}, R_{\rm out}) = (150 r_{\rm S}, 250 r_{\rm S})$ for $z = 230$--$250 r_{\rm S}$, and
$(R_{\rm in}, R_{\rm out}) = (300 r_{\rm S}, 400 r_{\rm S})$ for $z = 380 r_{\rm S}$,
respectively.

The ACFs show several noteworthy features.
First, no strong correlation is found at low altitudes, $z \ltsim 200 r_{\rm S}$,
meaning that no clumps there are.
Second, we see a sharp peak in the ACFs for the data above $z \gtsim 250 r_{\rm S}$,
indicating that clumps are being formed at $\sim 250 r_{\rm S}$.
The typical clump size $\ell_{\rm cl}$ can be measured by the width of the central peak
and is $\sim 10 r_{\rm S}$.
Third, the shape of ACFs at higher $z \sim 250 r_{\rm S}$ do not change appreciable. 
This means, the clumps retain their shapes in the statistical sense even after moving upwards.
The clump size is 25 times the grid spacing ($\Delta r = \Delta z = 0.4 r_{\rm S}$).
We can thus conclude that each clump is well resolved in our simulations.

To understand what the typical size of $\sim 10 r_{\rm S}$ means physically,
we estimate the optical depth of clumps for this length scale
inserting typical clump density, $\rho_{\rm cl} \sim 10^{-6}$ g cm$^{-3}$, finding
\begin{equation}
\tau_{\rm cl} =  \kappa_{\rm es} \rho_{\rm cl} \ell_{\rm cl} \sim 1.
\label{tau_cl}
\end{equation}
That is, the clump size is on the order of one optical depth.
This is a very important property of the clumps to consider their formation mechanism.
Here, we note that the electron scattering opacity dominates over
the absorption (free-free and bound-free) opacities because of high gas temperature
in the present simulation; that is, $\chi/\rho \sim \kappa_{\rm es}$.
This relation quite generally holds in radiation-dominated accretion disks or flows.

We here integrated ACFs in the $R$-direction.
Note that clumps constitute an elongate shape along the outflow direction, 
which is obvious in the upper panels of Figure~\ref{fig:rho_contour}.
This feature is also important to understand their physical formation mechanism.

\subsection{Anti-correlation between matter density and radiation force}
To understand the physical cause of the clump formation,
we need to know the relationship between different physical quantities.
The fact that the typical clump size corresponds to one optical depth
indicates the relevant physical mechanism underlying the clump formation
to be somehow related to radiation processes.

\begin{figure}
\begin{center}
\FigureFile(85mm,55mm){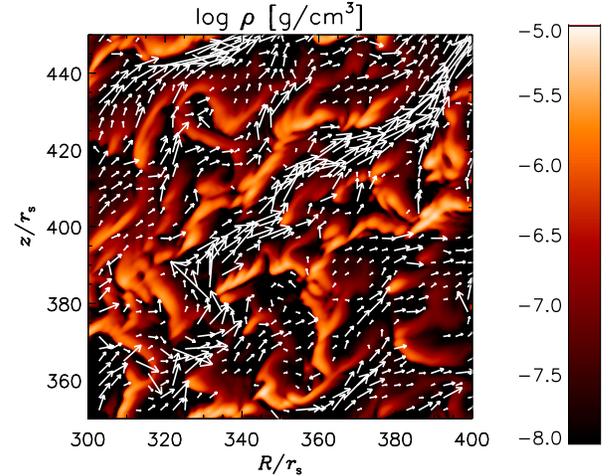}
\end{center}
\caption{
Color contour of matter density
overlaid with radiation force vector per unit mass (arrows) in the clumpy outflow region 
(corresponding to the region depicted in the right panels of Figure~\ref{fig:rho_contour}).
The arrows are displayed only in the region where their absolute values exceed
$5 \times {10}^{9}$ dyn g$^{-1}$.
}
\label{fig:rho-frad}
\end{figure}

Let us check the magnitude and the direction of the radiation force per unit mass, 
$ {\bm f}_{\rm rad} = \chi {\bm F}_0/\rho c$,
in the contour plot of the matter density, which is shown in Figure~\ref{fig:rho-frad}.
Anti-correlation between the matter density and the absolute magnitudes of the radiation force
is clear in the clumpy outflow region; i.e.,
we see longer arrows (representing stronger radiation force)
in the low-density regions (with dark colors).
It looks as if the radiative flux avoids dense (clumpy) regions and
instead selectively pass through low-density channels between the clumps.
The direction of the radiative flux seems to be responsible for the elongate shape of clumps.

\begin{figure}
\begin{center}
\FigureFile(85mm,55mm){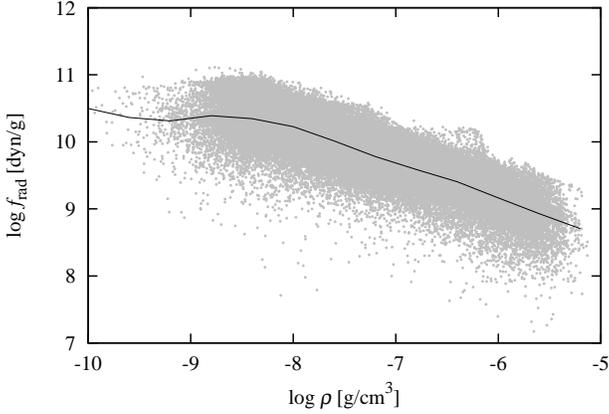}
\end{center}
\caption{
Correlation diagram between matter density and 
the absolute magnitudes of radiation force per unit mass in the clumpy outflow region
(corresponding to the area of the right panels of Figure~\ref{fig:rho_contour}).
}
\label{fig:diag}
\end{figure}

To see this anti-correlation nature more explicitly,
we plot in Figure~\ref{fig:diag} the values of matter density and 
radiation force in each grid point in the clumpy outflow region.
Again, the anti-correlation is clear.
Moreover, the relationships between
theses two quantities are roughly expressed by a power-law,
\begin{equation}
f_{\rm rad} \propto \rho^{-a},
\end{equation}
where a power-law index is found to be $a \sim 0.5$ at
$\rho \gtsim 10^{-8}$ g cm$^{-3}$ while $a \sim 0$ at $\rho \ltsim 10^{-8}$ g cm$^{-3}$.
The break of the power-law index is understood in terms of the optical depth.
In the optically thick media, where the radiative diffusion approximation holds,
the radiation flux is inversely proportional to
the optical depth (and, hence, the matter density); i.e.,
\begin{equation}
{f}_{\rm rad} \propto {F_0} \propto \frac{E_0}{\rho}.
\label{thick}
\end{equation}
In the optically thin media where the streaming limit applies,
on the other hand, the absolute value of the radiation force is given by
\begin{equation}
{f}_{\rm rad} \propto {F_0} \propto E_0.
\label{thin}
\end{equation}
We thus conclude that the break point of the power-law index
indicates the boundary between optically thick and thin limit.
As long as the radiation energy density $E_0$ is 
more or less homogeneous, which is actually the case,
the radiation force is weaker in the optically thick medium,
roughly inversely proportional to the optical depth, $f_{\rm rad} \propto \tau^{-1}$.
In the perturbed radiation-supported atmosphere,
the radiative flux tends to avoid high-density regions.

\begin{figure}
\begin{center}
\FigureFile(85mm,55mm){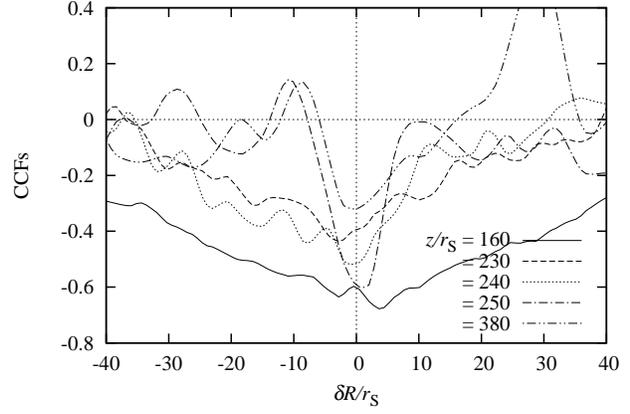}
\end{center}
\caption{
Cross-correlation functions (CCFs) between the matter density
and the absolute magnitudes of radiation force per unit mass in the $R$-direction ($\delta R$)
as a function of the typical height at the elapsed time of $t = 9$ s.
The negative value at the zero separation means the anti-correlation between each quantity.
}
\label{fig:cross}
\end{figure}

Let us illustrate the anti-correlation in a somewhat different way.
Figure~\ref{fig:cross} shows the cross-correlation functions (CCFs)
between the matter density and the absolute magnitudes of radiation force per unit mass.
The integration range is the same as Figure~\ref{fig:auto}. 
Anti-correlation is expressed by the negative value at zero separation.
We also notice that these anti-correlations actually evolve as
the flow moves upward (towards higher $z$).

\subsection{Structural Variations of a Gas Element}
\begin{table*}
 \begin{center}
 \caption{Characteristic timescales in the clumpy outflow.}
 \label{table:timescale}
  \begin{tabular}{llc}
   \hline
   \multicolumn{1}{c}{Various timescales}   &  Definitions & Values  \\
   \hline
    Dynamical time scale of outflow & $200 r_{\rm S}/v$   & $\gtsim 10^{-1}$ s \\
    Sound crossing time scale  & $\ell_{\rm cl}/c_{\rm s} = \ell_{\rm cl} (p_{\rm gas}/\rho)^{-1/2}$   & $\gtsim 10^{-1}$ s \\
    Joule heating time scale  & $e/Q^{+}_{\rm Joule} = e c^2 / 4 \pi \eta J^2$   & $\gtsim 10^{-2}$ s \\
    Radiative heating time scale  & $e/Q^{+}_{\rm rad} = e / c \chi_{\rm abs} E_0$   & $\gtsim 10^{-4}$ s \\
    Radiative cooling time scale  & $e/Q^{-}_{\rm rad} = e / 4 \pi \chi_{\rm abs} B_{\rm bb}$   & $\gtsim 10^{-4}$ s \\
   \hline
  \end{tabular}
\end{center}
\end{table*}

In order to understand how clumps form and grow at $z \gtsim 250 r_{\rm S}$,
we pick up one test gas element, follow its trajectory, 
and plot the time evolution of its physical quantities.
The selected gas elements travel from $(R,z)=(94 r_{\rm S}, 150 r_{\rm S})$
(within the accretion flow) at the elapsed time of $t = 8.8$ s
and reaches the point $(R,z)=(390 r_{\rm S}, 480 r_{\rm S})$ at $t = 9.7$ s.
The trajectory of the element is calculated by the following equation,
\begin{equation}
\bm{r}^{n+1} = \bm{r}^n + \bm{v}^n \Delta t,
\end{equation}
where $\bm{r}^n (n=1, 2, \cdots)$ is the position vector of the test particle
at the $n$-th time step ($t = t^n$), $\bm{v}^n$ is the velocity vector at $t = t^n$ 
obtained from the simulation data, and $\Delta t =t^{n+1}-t^n$ is the time interval.
We set the time interval to be $\Delta t = 10^{-4}$, 
which is shorter than any physical time scales (Table~\ref{table:timescale}).

\begin{figure}
\begin{center}
\FigureFile(85mm,55mm){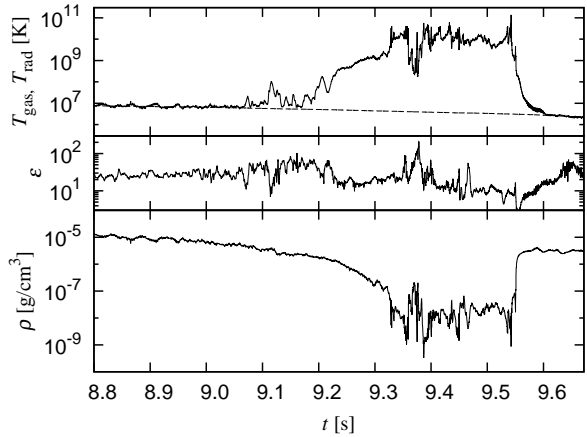}
\end{center}
\caption{
Time variations of some physical quantities of one test clump element.  
Top panel: the gas temperature (solid lines) and the radiation temperature (dashed line). 
Middle panel: the ratio of the radiation energy density to the sum of the gas and
magnetic energy densities. Bottom panel: the matter density.
This gas element emerging from the accretion flow at the elapsed time of $t=8.8$ s,
first moves through a low density, high temperature inter-clump region region at $t \sim 9.2$--$9.5$ s,
and merges into a high density, low temperature clump at $t \sim 9.6$ s.
}
\label{fig:traj}
\end{figure}

Figure~\ref{fig:traj} show a typical time evolution of gas elements.
Here, the top, middle, and bottom panels indicate gas and radiation temperatures, 
the ratio of the radiation energy density to the sum of the gas and
magnetic energy densities expressed by $\varepsilon = E_0/(e+B^2/8\pi)$, 
and the matter density, respectively. 
We find that most gas elements go through the low gas density and the high
gas temperature region before forming the clump.
Initially (at $t \ltsim 9.2$ s), this gas element was within the dense accretion flow,
in which matter and radiation is strongly coupled; i.e., $T_{\rm gas} = T_{\rm rad}$.
At $t \sim 9.2$--$9.5$ s, the decoupling occurs as the matter density decreases.
Finally, the hot gas element merges with pre-existed cold clump and cools down
($\rho \sim 10^{-6}$ g cm$^{-3}$; $T_{\rm gas} \sim 10^{6}$ K).

The gas temperature in the hot region (inter-clump space) is much larger than the
radiation temperature due probably to compressional heating caused by weak shock. 
We have confirmed
that kinematic energy density is comparable or slightly exceeds the thermal energy of gas.
The coupling between the radiation and matter
is so weak there that the radiative cooling cannot be effective.
(This contrasts with the case of clump region, in which
tight coupling between the radiation and matter leads to the equal temperatures.) 
Further,  Joule heating can assist the energy gain
but cannot be a main heating source,
since we will show in section 4.2 that
similar high temperatures are obtained by radiation-MHD
simulations with no magnetic fields. 
The gas entropy is much higher in the hot region than that within the clumps.
Therefore, the hot region is buoyant. 
Cautions should be taken here, however, since inverse-Compton scattering is
not considered in the present simulations. We expect somewhat lower temperature
in the inter-blob space (\cite{Kaw+09}).

To see what force is responsible for the acceleration of the clumpy outflow,
we compare the time-averaged forces over $t = 9.60$--$9.65$ s
by using the data in Figure~\ref{fig:traj} (see Table~\ref{table:value}).
Here, 
$f_{\rm cen} = v_\varphi^2R^{-1}$ is the centrifugal force,
$\bm{f}_{\rm Lor} = - \rho^{-1} \nabla p_{\rm mag} + \rho^{-1} \nabla \cdot (\bm{BB}/4\pi)$ is 
the Lorenz force (the magnetic-pressure force plus the tension force),
$\bm{f}_{\rm gas} = - \rho^{-1} \nabla p_{\rm gas} $ is the gas-pressure force,
$\cos \theta_{\rm cl} = v_R (v_R^2+v_z^2)^{-1/2}$ and 
$\sin \theta_{\rm cl} = v_z (v_R^2+v_z^2)^{-1/2} $
are the radial and vertical components of the acceleration of the gas particle, respectively.
We understand that the continuum radiation force is the driving force of the clumpy outflow.
This result is reasonable, since the clumpy outflow region is totally radiation-energy dominated
(see the middle panel in Figure~\ref{fig:traj}).

\begin{table*}
 \begin{center}
 \caption{
 Comparison of the forces asserted on the clumpy outflow.}
 \label{table:value}
  \begin{tabular}{llc}
   \hline
   \multicolumn{1}{c}{Various forces\footnotemark[$*$]}   &  Definitions & Values  \\
   \hline
    Continumm radiation force  & $f_{\rm rad}^{\rm para} = f_{\rm rad}^R\cos{\theta_{\rm cl}} + f_{\rm rad}^z\sin{\theta_{\rm cl}}$   & $\sim 3 \times 10^8$ dyn g$^{-1}$ \\
    Centrifugal force  & $f_{\rm cen}^{\rm para} = f_{\rm cen}\cos{\theta_{\rm cl}}$   & $\sim 1 \times 10^8$ dyn g$^{-1}$ \\
    Lorenz force & $f_{\rm Lor}^{\rm para} = f_{\rm Lor}^R\cos{\theta_{\rm cl}} + f_{\rm Lor}^z\sin{\theta_{\rm cl}}$   & $\sim 3 \times 10^5$ dyn g$^{-1}$ \\
    Gas pressure force & $f_{\rm gas}^{\rm para} = f_{\rm gas}^R\cos{\theta_{\rm cl}} + f_{\rm gas}^z\sin{\theta_{\rm cl}}$   & $\sim 3 \times 10^4$ dyn g$^{-1}$ \\
   \hline
   \multicolumn{3}{@{}l@{}}{\hbox to 0pt{\parbox{115mm}{\footnotesize
       \par\noindent
       \footnotemark[$*$] The radial and vertical components of the acceleration of the gas elements are 
       expressed by $\cos \theta_{\rm cl} = v_R (v_R^2+v_z^2)^{-1/2}$ and 
       $\sin \theta_{\rm cl} = v_z (v_R^2+v_z^2)^{-1/2} $, respectively.
    }\hss}}
  \end{tabular}
\end{center}
\end{table*}

\section{Discussion}
\subsection{Mechanism of creating clumpy outflow}
As we have seen in the previous section,
the clumpy outflow has a number of unique features:
(1) clumpy structure appears in the layer
where upward radiation force overcomes downward gravity force,
(2) a clump size is about one optical depth,
(3) there is a clear anti-correlation between the matter density and radiation force,
(4) temperature variations of some clumps are not monotonic increase nor monotonic decrease.
On the basis of these facts we discuss plausible physical mechanism of clump outflow.

The fact (4) obviously indicates that clump formation cannot be explained by a thermal instability,
which cause that the gas temperature monotonically decreases (increases)
with a monotonic increase (decrease) of the matter density (e.g., \cite{BalSok89}).
The cooling rate is expressed as
$ Q^-_{\rm rad} \propto \rho^2 T_{\rm gas}^{1/2} \propto T_{\rm gas}^{-3/2}$
under the condition of constant gas-pressure, 
which is a good assumption in the present simulations (see Fig. \ref{fig:traj}).
Since the thermal instability criterion is written as (e.g. \cite{Kat+08})
$  {d Q^+_{\rm rad}}/{d T_{\rm gas}} > {d Q^-_{\rm rad}}/{d T_{\rm gas}}$,
the system should be unstable, if the heating rate $Q^+_{\rm rad}$ does not depend on temperature.
This is the case corresponding to the interstellar medium.
In the present case, we evaluate 
$ Q^+_{\rm rad} \propto (\kappa_{\rm ff} + \kappa_{\rm bf}) \rho E_0 \propto T^{-11/2}_{\rm gas}$
for the Kramers-type opacity ($\kappa_{\rm ff} + \kappa_{\rm bf} \propto \rho T^{-7/2}_{\rm gas}$) 
adopted in the present study and under the conditions of constant gas-pressure 
and constant radiation-energy density ($E_0$).
Then, the system should be thermally stable.
However, it is not easy to express the heating rate in such simple forms
for such dynamically evolving media as those we study, 
since compressional heating may play a key role.

The fact (1) means that the clumpy outflow is {\it Rayleigh-Taylor unstable}.
In luminous black hole inflow-outflow system,
radiation field acts like an effective gravitational field.
The green line of the upper left panel in Figure~\ref{fig:rho_contour} indicates that
the direction of the effective gravitational field is reversed by the strong radiation force of the flow.
Since the matter density decreases in the direction of the acceleration,
Rayleigh-Taylor instability should set out.
The overturning motion of gas elements will form seeds of clumpy density pattern.
We thus conclude that Rayleigh-Taylor instability is the primary cause of the clump formation.
\citet{JacKru11} performed linear stability analysis of 
a plane-parallel superposition of two media immersed in radiation field.
They examined the stability in the optically thin and thick limit, 
finding that both cases could be Rayleigh-Taylor unstable.
Such radiation-induced Rayleigh-Taylor instability also appears
in massive star formation system and H\emissiontype{II} region
(see also \cite{MatBlu+77}; \cite{Kru+09}).

The Rayleigh-Taylor instability, however, cannot explain the facts (2) and (3), 
which rather imply radiation processes being somehow involved.
We thus further examine radiation-related instabilities.
From Equation (\ref{thick}) and (\ref{thin}),
radiation force could work as the positive feedback to grow the initial perturbation of the matter density, 
leading to a formation of inhomogeneous density pattern.
\citet{Sha01a} made a global linear stability analysis for the optically thick,
radiation-dominated atmospheres of the super-Eddinton outflow from stars.
He has clearly shown that perturbations with wavelengths on the order of
pressure scale-height $H$ grow on the order of sound crossing timescales under a fixed 
radiation-temperature condition at the bottom.
This is a radiation-hydrodynamic instability and is expected to 
create inhomogeneous (porous) density structure just below photosphere of the objets 
(see also \cite{Sha01b}; \cite{Fuk03}).
Note that one pressure scale-height roughly corresponds to one optical depth 
just below the photosphere.
That is, this instability may help forming clumps of one optical depth in the present study.

We here found the anti-correlation between the matter density and 
the absolute value of radiation force per unit mass in the present analysis.
In contrast, \citet{Sha01a} claimed anti-correlation between matter density 
and radiation energy density, which we do not confirm in our simulations. 
The reason for this should be clarified in a future work.

In this simulation, we assumed that the flow is axisymmetric with respect to the rotation axis; i.e.,
the dense outflowing gases in the present simulation are not clumpy but annular.
However, since the Rayleigh-Taylor and the radiation-hydrodynamic instability
would work the clumpy structure formation even in the three-dimensional simulations.

\subsection{Role of Magnetic Field in Clump Formation}
\begin{figure}
\begin{center}
\FigureFile(85mm,55mm){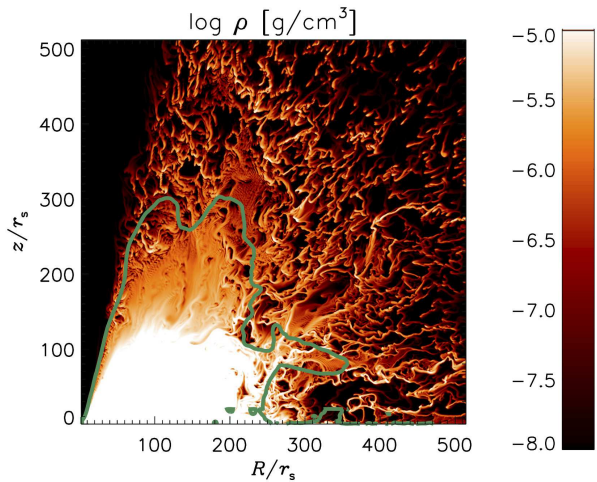}
\end{center}
\caption{
Two-dimensional structure of a supercritical accretion flow and the associated outflow
simulated by the same radiation-MHD code but with no magnetic field.
Green lines indicate the surface which upward radiation force equals 
downward gravity force of the central black hole.
Similar clumpy outflow structure to that found in Figure~\ref{fig:rho_contour} is confirmed.
}
\label{fig:non-mag}
\end{figure}

When strong, ordered magnetic field are present in radiation-dominated atmosphere,
the magnetic photon bubble instability may also work to produce inhomogeneous structure
on the condition of $F_{\rm 0} \gtsim E_{\rm 0} c_{\rm s}$ 
(\cite{Aro92}; \cite{Gam98}; \cite{BlaSoc03}).
\citet{Tur+05} performed the local radiation-MHD simulations
of the standard thin (Shakura-Sunyaev) disk
with the initial mass accretion rate 10\% of the Eddington limit for a 10\% radiative efficiency,
and reported the occurrence of the magnetic photon bubble instability, giving rise to 
inhomogeneous density structure.
They also assert that the radiation-hydrodynamic instability of \citet{Sha01a} does not occur,
although the this instability grows faster than the magnetic photon bubble instability,
since the fixed radiation temperature lower boundary is not satisfied.

To see if the existence of magnetic field is essential for the clumpy structure formation in our simulations,
we performed {\it non-magnetic} radiation-hydrodynamic simulations.
In this simulation, we adopted the data of the radiation-MHD calculation
at the elapsed time of 7 s as the initial condition
and then run the same code but by artificially vanishing magnetic field everywhere.
Once magnetic field are made zero, they never appear forever
(see induction equation in Equation \ref{ind}).
We should also note that there is no accretion motion in this simulation,
since the disk viscosity, which is of a magnetic origin, disappears.
This simulation is nevertheless useful to see the dynamical properties of the non-magnetized outflow
which takes place on much shorter timescales than the viscous timescale of the underlying disk.

Figure~\ref{fig:non-mag} illustrates the resultant density distribution in the whole computational domain
at the elapsed time of $t = 9$ s. 
The green lines indicate the surface of 
$\chi F_0/c =  \rho |\nabla \psi_{\rm PN}|$. 
It clearly shows again clumpy outflow structure above $z \gtsim 250 r_{\rm s}$,
where the upward radiation force overcomes the downward gravity force.
We further confirm the typical clump size to be $10 r_{\rm S}$ from the ACF analysis and
anti-correlation between the matter density and the radiation force in the non-magnetic case, as well.
We can thus conclude that magnetic field is not essential for creating clumpy outflows. 
In other words, the magnetic photon bubble instability is not a primary cause of clump formation.
This result, however, 
does not exclude a possible occurrence of magnetic photon bubble instability in a real situation, 
since the wavelength with the fastest growth rate is $\sim (3p_{\rm gas}/E_0)H \ll H$,
which is too short to resolve by the present simulations.
We need finer grid spacing, than the present case, 
to see the effects of magnetic photon bubble instability, if it really occurs.

\subsection{Comparison with the Observational facts}
As was mentioned in Introduction, 
the emergence of outflow from various types of black hole objects has been reported
through a number of X-ray observations.
The clumpy nature of the outflow has also been suggested recently by the observations of
luminous accretion flows, such as ultra-luminous X-ray sources (ULXs) and bright AGNs.
That is, the clumpy outflow nature does not depend on black hole masses observationally. 
We note that our results of the inflow-outflow structure also apply to the cases with supermassive black holes, 
since Thomson scattering dominates over absorption opacities in the supercritical accretion regimes, 
whatever black hole masses will be.
The matter density decreases with an increase of the black hole mass, $\rho \propto M^{-1}$, 
for a fixed mass accretion rate.
Hence, the radiation force on the Thomson scattering ($\propto \rho^0$) is
dominant over the bound-free and free-free absorptions ($\propto \rho$) 
even in the case of massive holes.

\begin{figure}
\begin{center}
\FigureFile(85mm,55mm){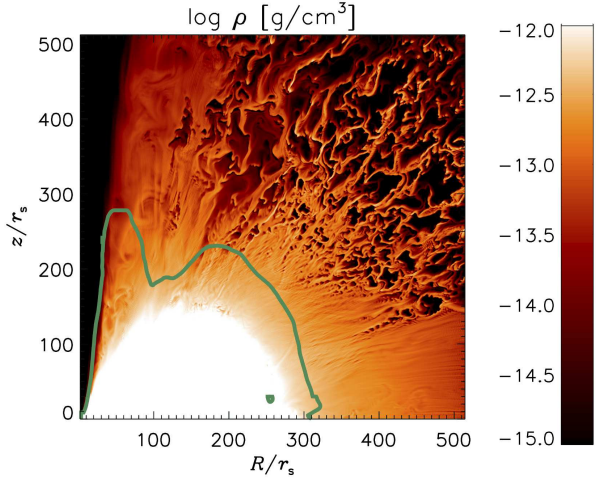}
\end{center}
\caption{
Two-dimensional structure of a supercritical accretion flow and the associated outflow
around a black hole with $M = 10^8 M_\odot$ at the elapsed time of $t = 7 \times 10^{7}$ s.
Green lines indicate the surface which upward radiation force equals 
downward gravity force of the central black hole.
The clumpy structure does not depend on the black hole mass.
}
\label{fig:AGN}
\end{figure}

We actually performed the same radiation-MHD simulations for the case 
with a supermassive black hole ($M = 10^8 M_\odot$).
The resultant density contours shown in Figure~\ref{fig:AGN}
clearly demonstrate the emergence of similar clumpy outflow patterns.
By calculating ACFs, we also confirmed that the typical clump size
is $\sim 10 r_{\rm S}$, corresponding to 
$\tau_{\rm cl} = \kappa_{\rm es} \rho_{\rm cl} \ell_{\rm cl} \sim 1$
(since $\rho_{\rm cl} \propto M^{-1}$ while $\ell_{\rm cl}\propto M$).

Based on the variability study of the ULX, NGC~5408~X-1,
\citet{Mid+11} found that it has a similar spectra with
the black hole binary (BHB), GRS~1915+105 and some extreme Narrow-Line Seyfert 1s.
They conclude that the underlying accretion flow should be supercritical
and the associated outflow has clumpy structure
with the variability on time-scales of several tens of seconds (see also \cite{Mid+12}).
\citet{Tom+10} detected blueshifted Fe K absorption lines from several AGNs
whose the bolometric luminosities are estimated to be close to the Eddington one,
and suggested the presence of outflows from
the central regions with mildly relativistic velocities, in the range $0.04$--$0.15c$:
so called Ultra-Fast Outflows (UFOs).
The ionization of the absorbers is in the range $\xi \sim 10^3$--$10^6$ erg cm s$^{-1}$, 
and the column density is in the interval $N_{\rm H} \sim 10^{22}$--$10^{24}$ cm$^{-2}$ (\cite{Tom+11}).
The observed spectral variability on time-scales of $\sim$ days (e.g., \cite{Bra+07}; \cite{Cap+09})
suggests that the outflows have clumpy structure (\cite{Tom+12}).

Let us compare our results with these observational facts.
It is important to note that once a clump passes across our line of sight, 
it will produce significant absorption, since the clump optical depth is about unity.
To see how frequently such obscuration occurs, 
we calculate the spatial covering factor of the clumps, $\mathcal{C}$,
by integrating the fraction of sky covered by each clump element.
The factor can be derived by
\begin{equation}
\mathcal{C} = 2 \int^{\infty}_{r_{\rm ph}} \int^{\theta_2}_{\theta_1} \int^{2\pi}_{0} n_{\rm cl}
\left( \frac{ \pi \ell_{\rm cl}^2 } {16\pi r^2} \right) r^2 \sin \theta dr d\theta d\varphi,
\label{cover}
\end{equation}
where $n_{\rm cl}$ is the number density of the clump,
$r_{\rm ph}$ is the location of the photosphere,
and $(\theta_2-\theta_1)$ is the opening angle of clumpy outflow, respectively  (e.g., \cite{BotFer01}).
The spherical polar coordinates ($r$, $\theta$, $\varphi$) are used.
Note that the clump size $\ell_{\rm cl}$ is not radius but diameter.
The number density is given by
\begin{equation}
n_{\rm cl} \sim \frac{\dot{M}_{\rm out}}{r^2 m_{\rm cl} v_r \Omega},
\end{equation}
where we use the following relation:
the mass outflow rate $\dot{M}_{\rm out} = r^2 \rho v_r \Omega$;
the matter density of the outflow $\rho \sim m_{\rm cl} n_{\rm cl}$;
the unit mass of the clump $m_{\rm cl} = \pi \ell_{\rm cl}^3 \rho_{\rm cl}/6$;
the solid angle of the outflow $\Omega = 2 \int^{\theta_2}_{\theta_1} \int^{2\pi}_{0} \sin \theta d\theta d\varphi$.
Using Equation (\ref{tau_cl}),
the covering factor is estimated to be
\begin{eqnarray}
\mathcal{C} \sim 0.3 \tau_{\rm cl}^{-1} 
\Bigg( \frac{\dot{M}_{\rm out}}{10 L_{\rm E}/c^2} \Bigg)
\Bigg( \frac{v_r}{0.1c} \Bigg)^{-1} 
\Bigg(\frac{r_{\rm ph}}{250 r_{\rm S}} \Bigg)^{-1}.
\label{cover_deri}
\end{eqnarray}
Note that the covering factor does not depend on the black hole mass nor the size of the clump,
as long as the clump size is expressed in terms of the Schwarzschild radius.
We thus expect occasional obscuration of the light from the center by clumpy outflow, 
being thus in agreement with the observations.

Next, we estimate the photoionization parameter of clumps.
The photoionization parameter, $\xi$, is first proposed by \citet{Tar+69}
and is expressed by
\begin{equation}
\xi = \frac{L_{\rm X}}{n_{\rm cl} r^2} ,
\end{equation}
where $L_{\rm X}$ is the luminosity in the X-ray band.
Using the relation, $\tau_{\rm cl} = n_{\rm cl} \sigma_{\rm T} \ell_{\rm cl}$,
the photoionization parameter is estimated to be
\begin{eqnarray}
\xi \sim && 10^3 \tau_{\rm cl}^{-1}  \nonumber \\
&& \left( \frac{L_{\rm X}}{0.1 L_{\rm E}} \right)  
 \left( \frac{\ell_{\rm cl}}{10 r_{\rm S}} \right) \left( \frac{r}{500 r_{\rm S}} \right)^{-2} \; {\rm erg\;cm\;s^{-1}}.
\label{equ:photo}
\end{eqnarray}
Equation (\ref{equ:photo}) indicates 
that clumpy outflows are mildly ionized (\cite{KalMcC82}).
Although the parameter does not explicitly depend on the black hole mass,
strictly speaking, 
the luminosity in the X-ray band $L_{\rm X}$ depends on the black hole mass;
$L_{\rm X} \sim L_{\rm bol}$ for Galactic sources and $L_{\rm X} \sim 0.1 L_{\rm bol}$ for AGNs.
Here, $L_{\rm bol}$ is the bolometric luminosity of the source.
We assume $L_{\rm bol} \sim L_{\rm E}$ in Equation (\ref{equ:photo}).
We thus expect line absorption features to be observed, in agreement with the observations.

Finally, let us estimate the variability timescale of the clumpy outflow $t_{\rm cl}$,
which is given by
\begin{eqnarray}
t_{\rm cl} = \frac{\ell_{\rm cl}}{v_\varphi} \sim \frac{\ell_{\rm cl}r}{\sqrt{GMr_{\rm ph}}}.
\end{eqnarray}
We here assume that 
the angular momentum, which was $\sqrt{GM r_{\rm ph}}$ at the photosphere (at $r=r_{\rm ph}$), 
is conserved even when it travels to the radius, $r$.
We thus estimate 
\begin{eqnarray}
t_{\rm cl} \sim 6 \left( \frac{M}{10^8 M_\odot} \right)  \left( \frac{r_{\rm ph}}{250 r_{\rm S}} \right)^{-\frac{1}{2}} \left( \frac{r}{500 r_{\rm S}} \right) \; {\rm day},
\label{equ:vari}
\end{eqnarray}
in good agreement with the observations of AGN.
This estimation cannot explain the long variation timescales of the ULXs, 
on the order of several tens of seconds, unless the black hole is large (\cite{Mid+11}). 
If the same mechanism applies to ULXs, they should contain supermassive black holes.

The simulation results can account for the basic properties of the UFOs of AGN observed with X-ray.
We should note, however, that we adopt gray approximation in the present simulations, 
meaning that we cannot properly discuss the spectral energy distribution (SED). 
More detailed comparison with the observations, in terms of absorption spectra, is left for future issue.

\subsection{The Origin of Broad-Line Clouds?}
The broad-line region (BLR) is a standard AGN ingredient (e.g. \cite{Fer+92}; \cite{Pet97}): 
it is located inside the obscuring torus, above and below the accretion flow, 
and at a distance of parsec scale from the supermassive black hole. 
Can we explain BLR clouds by clumpy outflow?
From simulation data,
the electron number density and the gas temperature of the clumps are estimated 
$\sim \rho_{\rm cl}/m_{\rm p} \sim 10^{11}$ cm$^{-3}$
and $\sim 10^4$ K for $M = 10^8 M_\odot$,
since the gas temperature is proportional to $M^{-1/4}$
for radiation-dominated atomosphere.
The volume filling factor is expressed by
\begin{equation}
\mathcal{F} = \frac{N_{\rm cl} ( \pi \ell_{\rm cl}^3/6)}{(4\pi r_{\rm BLR}^3/3)}
= \frac{N_{\rm cl} \ell_{\rm cl}^3}{8r_{\rm BLR}^3},
\label{fill}
\end{equation}
where $N_{\rm cl} = \iiint n_{\rm cl} r^2 \sin \theta dr d\theta d\varphi$ 
is the number of clumps, 
and $r_{\rm BLR}$ is the size of the BLR.
By using some typical values, we got the filling factor  
\begin{eqnarray}
\mathcal{F} \sim && 10^{-7}  \nonumber \\
&&  \Bigg( \frac{\dot{M}_{\rm out}}{10 L_{\rm E}/c^2} \Bigg) 
 \Bigg( \frac{v_r}{0.1c} \Bigg)^{-1} 
 \Bigg(\frac{\ell_{\rm cl}}{10 r_{\rm S}} \Bigg)
 \Bigg(\frac{r_{\rm BLR}}{10^5 r_{\rm S}} \Bigg)^{-2}.
\label{fill_deri}
\end{eqnarray}
This value is consistent with the observational values of the BLR clouds (\cite{Pet97}).
In this estimation, we assumed that each clump keeps its form up to the distance of 
$r_{\rm BLR} = 10^5 r_{\rm S} \sim 1 {\rm pc}$.
We check if clumps survive by comparing various timescales.
The gas temperature of the clumps hardly change due to the matter-radiation coupling occur in them,
whereas that of the ambient media would remain to be hot 
since the radiative cooling cannot be effective.
The thermal conduction time scale, 
which is expressed by $3\kappa_{\rm Sp}^{-1}n_{\rm cl}k_{\rm B}\ell_{\rm cl}^2T^{-5/2}$,
is also much lager than the dynamical time scale,
where $\kappa_{\rm Sp}$ is the Spitzer conductivity coefficient.
Thus the clumps are expected to remain to large domain.
To examine the stability of the clumps,
the study with more larger simulation box is needed.

The estimated typical values from the super-Eddington outflow 
with the mass outflow rate of $\dot{M}_{\rm out} = 10 L_{\rm E}/c^2$ 
is in agreement with the observational facts.
Importantly, the photon luminosity of the underlying accretion flow 
is $\sim L_{\rm E}$ in spite of the large mass accretion and outflow rate,
since the a large amount of photons inside the accretion flow is trapped and swallowd 
to the central black hole (so-called the photon-tapping effect) (\cite{Ohs+05}).

\citet{Eli12} attempts to explain the observations of type~I and II AGNs
based on the assumption that clumpy outflow gas originating from the accretion flow
is distributed around luminous AGN. 
His model can nicely explain the observed BLR disappearance at low luminosity (\cite{EliHo09}).

\subsection{Possible line-driven outflow}
Although we only consider the continuum radiation force in the present study,
line force may also work in luminous black hole inflow-outflow system.
The line-driven force has been suggested to 
the one of the most promising mechanisms of AGN outflow,
since this can explain both acceleration and ionization states of the outflow materials (\cite{Pro+00}).
When matter absorbs UV photons, it gets momentum from them,
and is accelerated towards the direction opposite to that the photons come from.
The ionized X-ray photon distribution determines the outflow structure
(\cite{Cas+75}; \cite{RisElv10}; \cite{Nom+12}).
Since the SED of the underlying accretion flow strongly 
depends on the dynamics of line-driven outflows,
it is unclear whether the line-driven mechanism does really work or not in the present study
(note that we assume the gray approximation for radiation transfer).
\citet{KurPro09} performed the three-dimensional hydrodynamical simulations of line-driven wind, 
reporting the clumpy outflow structure due to the rotation shear.
Their finding is, however, independent of our work, in which radiation-hydronamical
effects seem to play a critical role.
The performance of the global radiation-MHD simulation of the line-driven outflow is future work.

\section{Conclusion}
In this paper, we performed the 
two-dimensional global radiation-MHD simulations of supercritical accretion flows
onto black holes in the larger computational box than our previous one
and examined the properties of the outflow from the accretion flow in details.
Here are our new findings:
\begin{itemize}
\item 
The outflows associated with supercritical (or super-Eddington) accretion flows 
have a clumpy structure above heights of $\sim 250 r_{\rm S}$.
The typical clump size is $\sim 10 r_{\rm S}$, which corresponding to about one optical depth,
and their shapes are slightly elongated along the outflow direction.
In the clumpy outflow region, a clear anti-correlation is seen 
between matter density and the absolute value of radiation force per unit mass.
\item 
Rayleigh-Taylor instability is the most plausible cause of the clump formation,
since the clumpy structure appears in the layer 
where upward radiation force overcomes downward gravity force.
In addition, a radiation-hydrodynamic instability, 
which arises when radiation funnels through radiation-pressure supported atmosphere, 
may also help forming clumps of one optical depth.
Magnetic filed is not essential for this clumpy structure formation. 
\item 
The spatial covering factor of the clumps are estimated $\sim 0.3$ 
for typical parameters of supercritical accretion flow, 
regardless of the black hole mass nor the size of the clump.
This mean that presence of the occasional obscuration of the light from the center by clumpy outflow.
For the clumpy outflows from supermassive black hole with $M = 10^8 M_\odot$,
the photoionization parameter and the variability timescale of the clump 
are estimated $\sim 10^3$ erg cm s$^{-1}$ and $\sim 6$ day, 
in good agreement with the observational results.
\item
If the clumps remain to parsec scale, the volume filling factor is estimated $\sim 10^{-7}$,
and the clumpy outflow are consistent with the observational values of BLR clouds.
\end{itemize}

\bigskip
The authors would like to thank the anonymous referee for
important comments and suggestions.
This work is supported in part by the Grant-in-Aid of
Ministry of Education, Culture, Sports, Science, and Technology (MEXT) (22340045, SM; 24740127, KO)
and by the Grant-in-Aid for the global COE programs on The Next Generation of Physics, Spun from Diversity
and Emergence from MEXT (SM). Numerical computations were in part carried out on Cray XT4 at Center for
Computational Astrophysics (CfCA) of National Astronomical Observatory of Japan.

\appendix
\section*{Correlation Function}
The correlation function is a measure of similarity of two waveforms as a function of a lag.
Correlation analysis is used to find periodic patterns and/or coherent lengths in data.
The correlation function $f_{\rm corr}(L)$ of two sample populations $x$ and $y$ as a function of the lag $L$ is calculated by,
\[
  f_{\rm corr}(L)  = \left\{ \begin{array}{ll}
    \frac{ {\displaystyle \sum_{k=0}^{N-L-1} } (x_{k+L}-\bar{x}) (y_{k}-\bar{y})}
                  { \sqrt{\left[
                    {\displaystyle \sum_{k=0}^{N-1} }     (x_{k}-\bar{x})^2  \right]
                             \left[
                    {\displaystyle \sum_{k=0}^{N-1} }     (y_{k}-\bar{y})^2  \right]}
                   }
     & \rm{for} \; \it{L} \rm{< 0} \\
     \frac{ {\displaystyle \sum_{k=0}^{N-L-1} } (x_{k}-\bar{x}) (y_{k+L}-\bar{y})}
                  { \sqrt{\left[
                    {\displaystyle \sum_{k=0}^{N-1} }     (x_{k}-\bar{x})^2  \right]
                             \left[
                    {\displaystyle \sum_{k=0}^{N-1} }     (y_{k}-\bar{y})^2  \right]}
                   }
     & \rm{for} \; \it{L} \rm{\ge 0}
  \end{array} \right.
\]
where $\bar{x}$ and $\bar{y}$ are the means of the sample populations $x = (x_0, x_1, x_2, ... , x_{N-1})$ and
$y = (y_0, y_1, y_2, ... , y_{N-1})$, respectively.
The correlation function is called auto-correlation function (ACF)
in the case of same waveform ($x = y$),
and cross-correlation function (CCF) 
in the case of different waveforms ($x \neq y$).


\begin{thebibliography}{}
\bibitem[Arons(1992)]{Aro92} Arons, J.\ 1992, \apj, 388, 561 
\bibitem[Balbus \& Soker(1989)]{BalSok89} Balbus, S.~A., \& Soker, N.\ 1989, \apj, 341, 611 
\bibitem[Blaes \& Socrates(2003)]{BlaSoc03} Blaes, O., \& Socrates, A.\ 2003, \apj, 596, 509 
\bibitem[Bottorff \& Ferland(2001)]{BotFer01} Bottorff, M., \& Ferland, G.\ 2001, \apj, 549, 118 
\bibitem[Braito et al.(2007)]{Bra+07} Braito, V., Reeves, J.~N., Dewangan, G.~C., et al.\ 2007, \apj, 670, 978 
\bibitem[Cappi et al.(2009)]{Cap+09} Cappi, M., Tombesi, F., Bianchi, S., et al.\ 2009, \aap, 504, 401 
\bibitem[Castor et al.(1975)]{Cas+75} Castor, J.~I., Abbott, D.~C., \& Klein, R.~I.\ 1975, \apj, 195, 157 
\bibitem[Di Matteo et al.(2005)]{DiM+05} Di Matteo, T., Springel, V., \& Hernquist, L.\ 2005, \nat, 433, 604
\bibitem[Elitzur \& Ho(2009)]{EliHo09} Elitzur, M., \& Ho, L.~C.\ 2009, \apjl, 701, L91 
\bibitem[Elitzur(2012)]{Eli12} Elitzur, M.\ 2012, \apjl, 747, L33
\bibitem[Ferland et al.(1992)]{Fer+92} Ferland, G.~J., Peterson, B.~M., Horne, K., Welsh, W.~F., \& Nahar, S.~N.\ 1992, \apj, 387, 95 
\bibitem[Fukue(2003)]{Fuk03} Fukue, J.\ 2003, \pasj, 55, 451 
\bibitem[Gammie(1998)]{Gam98} Gammie, C.~F.\ 1998, \mnras, 297, 929
\bibitem[Ganguly \& Brotherton(2008)]{GanBro08} Ganguly, R., \& Brotherton, M.~S.\ 2008, \apj, 672, 102
\bibitem[Gladstone et al.(2009)]{Gla+09} Gladstone, J.~C., Roberts, T.~P., \& Done, C.\ 2009, \mnras, 397, 1836 
\bibitem[Gonz{\'a}lez et al.(2007)]{Gon+07} Gonz{\'a}lez, M., Audit, E., \& Huynh, P.\ 2007, \aap, 464, 429 
\bibitem[Granato et al.(2004)]{Gra+04} Granato, G.~L., De Zotti, G., Silva, L., Bressan, A., \& Danese, L.\ 2004, \apj, 600, 580 
\bibitem[Jacquet \& Krumholz(2011)]{JacKru11} Jacquet, E., \& Krumholz, M.~R.\ 2011, \apj, 730, 116 
\bibitem[Kallman \& McCray(1982)]{KalMcC82} Kallman, T.~R., \& McCray, R.\ 1982, \apjs, 50, 263 
\bibitem[Kato et al.(2008)]{Kat+08} Kato, S., Fukue, J., \& Mineshige, S. 2008, Black-Hole Accretion Disks: Towards a New Paradigm (Kyoto: Kyoto University Press) 
\bibitem[Kawabata \& Mineshige(2009)]{KawMin09} Kawabata, R., \& Mineshige, S.\ 2009, \pasj, 61, 1135 
\bibitem[Kawashima et al.(2009)]{Kaw+09} Kawashima, T.,  Ohsuga, K., Mineshige, S., et al.\ 2009, \pasj, 61, 769 
\bibitem[Kawashima et al.(2012)]{Kaw+12} Kawashima, T., Ohsuga, K., Mineshige, S., et al.\ 2012, \apj, 752, 18 
\bibitem[Kotani et al.(2006)]{Kot+06} Kotani, T., Trushkin, S.~A., Valiullin, R., et al.\ 2006, \apj, 637, 486 
\bibitem[Krumholz et al.(2009)]{Kru+09} Krumholz, M.~R., Klein, R.~I., McKee, C.~F., Offner, S.~S.~R., \& Cunningham, A.~J.\ 2009, Science, 323, 754 
\bibitem[Kubota et al.(2007)]{Kub+07} Kubota, A., Dotani, T., Cottam, J., et al.\ 2007, \pasj, 59, 185 
\bibitem[Kurosawa \& Proga(2009)]{KurPro09} Kurosawa, R., \& Proga, D.\ 2009, \apj, 693, 1929 
\bibitem[Levermore \& Pomraning(1981)]{LevPom81} Levermore, C.~D., \& Pomraning, G.~C. 1981, \apj, 248, 321
\bibitem[Mathews \& Blumenthal(1977)]{MatBlu+77} Mathews, W.~G., \& Blumenthal, G.~R.\ 1977, \apj, 214, 10 
\bibitem[Middleton et al.(2011)]{Mid+11} Middleton, M.~J., Roberts, T.~P., Done, C., \& Jackson, F.~E.\ 2011, \mnras, 411, 644
\bibitem[Middleton et al.(2012)]{Mid+12} Middleton, M.~J., Sutton, A.~D., Roberts, T.~P., Jackson, F.~E., \& Done, C.\ 2012, \mnras, 420, 2969
\bibitem[Mihalas \& Weibel Mihalas(1984)]{MihMih84} Mihalas, D., \& Weibel Mihalas, B.\ 1984, New York: Oxford University Press, 1984,  
\bibitem[Miller et al.(2004)]{Mil+04} Miller, J.~M., Raymond, J., Fabian, A.~C., et al.\ 2004, \apj, 601, 450 
\bibitem[Miller et al.(2008)]{Mil+08} Miller, L., Turner, T.~J., \& Reeves, J.~N.\ 2008, \aap, 483, 437
\bibitem[Murray et al.(1995)]{Mur+95} Murray, N., Chiang, J., Grossman, S.~A., \& Voit, G.~M.\ 1995, \apj, 451, 498 
\bibitem[Neilsen et al.(2011)]{Nei+11} Neilsen, J., Remillard, R.~A., \& Lee, J.~C.\ 2011, \apj, 737, 69 
\bibitem[Neilsen et al.(2012)]{Nei+12} Neilsen, J., Petschek, A.~J., \& Lee, J.~C.\ 2012, \mnras, 421, 502
\bibitem[Nomura et al.(2012)]{Nom+12} Nomura, M., Ohsuga, K., Wada, K., Susa, H., \& Misawa, T.\ 2012, arXiv:1212.3075 
\bibitem[Ohsuga et al.(2005)]{Ohs+05} Ohsuga, K., Mori, M., Nakamoto, T., \& Mineshige, S.\ 2005, \apj, 628, 368
\bibitem[Ohsuga et al.(2009)]{Ohs+09} Ohsuga, K., Mineshige, S., Mori, M., \& Kato, Y. 2009, \pasj, 61, L7 (Paper I)
\bibitem[Ohsuga \& Mineshige(2011)]{OhsMin11} Ohsuga, K., \& Mineshige, S.\ 2011, \apj, 736, 2 (Paper II)
\bibitem[Paczy\'nski \& Wiita(1980)]{PacWii80} Paczy\'nski, B., \& Wiita, P.J. 1980, A\&A, 88, 23
\bibitem[Peterson(1997)]{Pet97} Peterson, B.~M.\ 1997, An introduction to active galactic nuclei (Cambridge: Cambridge University Press)
\bibitem[Ponti et al.(2012)]{Pon+12} Ponti, G., Fender, R.~P., Begelman, M.~C., et al.\ 2012, \mnras, 422, L11 
\bibitem[Pounds et al.(2003)]{Pou+03} Pounds, K.~A., Reeves, J.~N., King, A.~R., et al.\ 2003, \mnras, 345, 705 
\bibitem[Pounds \& Reeves(2009)]{PouRee09} Pounds, K.~A., \& Reeves, J.~N.\ 2009, \mnras, 397, 249 
\bibitem[Proga et al.(1998)]{Pro+98} Proga, D., Stone, J.~M., \& Drew, J.~E.\ 1998, \mnras, 295, 595
\bibitem[Proga et al.(2000)]{Pro+00} Proga, D., Stone, J.~M., \& Kallman, T.~R.\ 2000, \apj, 543, 686
\bibitem[Proga \& Kallman(2004)]{ProKal04} Proga, D., \& Kallman, T.~R.\ 2004, \apj, 616, 688 
\bibitem[Reeves et al.(2003)]{Ree+03} Reeves, J.~N., O'Brien,  P.~T., \& Ward, M.~J.\ 2003, \apjl, 593, L65 
\bibitem[Risaliti \& Elvis(2010)]{RisElv10} Risaliti, G., \& Elvis, M.\ 2010, \aap, 516, A89 
\bibitem[Shakura \& Sunyaev(1973)]{ShaSun73} Shakura, N. I., \& Sunyaev, R. A. 1973, A\&A, 24, 337
\bibitem[Shaviv(2001a)]{Sha01a} Shaviv, N.~J.\ 2001a, \apj, 549, 1093
\bibitem[Shaviv(2001b)]{Sha01b} Shaviv, N.~J.\ 2001b, \mnras, 326, 126
\bibitem[Stone \& Norman(1992)]{StoNor92} Stone, J.~M., \& Norman, M.~L.\ 1992, \apjs, 80, 753 
\bibitem[Takeuchi et al.(2009)]{Tak+09} Takeuchi, S., Mineshige, S., \& Ohsuga, K.\ 2009, \pasj, 61, 783
\bibitem[Takeuchi et al.(2010)]{Tak+10} Takeuchi, S., Ohsuga, K., \& Mineshige, S.\ 2010, \pasj, 62, L43
\bibitem[Tarter et al.(1969)]{Tar+69} Tarter, C.~B., Tucker, W.~H., \& Salpeter, E.~E.\ 1969, \apj, 156, 943 
\bibitem[Terashima \& Wilson(2001)]{TerWil01} Terashima, Y., \& Wilson, A.~S.\ 2001, \apj, 560, 139 
\bibitem[Tombesi et al.(2010)]{Tom+10} Tombesi, F., Sambruna, R.~M., Reeves, J.~N., et al.\ 2010, \apj, 719, 700 
\bibitem[Tombesi et al.(2011)]{Tom+11} Tombesi, F., Cappi, M., Reeves, J.~N., et al.\ 2011, \apj, 742, 44 
\bibitem[Tombesi et al.(2012)]{Tom+12} Tombesi, F., Cappi, M., Reeves, J.~N., \& Braito, V.\ 2012, \mnras, 422, L1 
\bibitem[Turner et al.(2005)]{Tur+05} Turner, N.~J., Blaes, O.~M., Socrates, A., Begelman, M.~C., \& Davis, S.~W.\ 2005, \apj, 624, 267 
\bibitem[Ueda et al.(2009)]{Ued+09} Ueda, Y., Yamaoka, K., \& Remillard, R.\ 2009, \apj, 695, 888
\end{thebibliography}
\end{document}